\documentclass[aps,pre,twocolumn]{revtex4-1}
\usepackage{amsmath}
\usepackage{amssymb}
\usepackage{graphicx}

\newcommand{\br}{{\bf r}}

\newcommand{\Tr}{\,{\rm{Tr}\,}}

\newcommand{\beqa}{\begin{eqnarray}}
\newcommand{\eeqa}{\end{eqnarray}}

\begin{document}
\title{Level density and level-spacing distributions of random, self-adjoint, non-Hermitian matrices} 
\author{Yogesh N. Joglekar}
\author{William A. Karr}
\affiliation{Department of Physics, 
Indiana University Purdue University Indianapolis (IUPUI), 
Indianapolis, Indiana 46202, USA}
\date{\today}
\begin{abstract}
 We investigate the level-density $\sigma(x)$ and level-spacing distribution $p(s)$ of random matrices $M=AF\neq M^{\dagger}$ where $F$ is a (diagonal) inner-product and $A$ is a random, real symmetric or complex Hermitian matrix with independent entries drawn from a probability distribution $q(x)$ with zero mean and finite higher moments. Although not Hermitian, the matrix $M$ is self-adjoint with respect to $F$ and thus has purely real eigenvalues. We find that the level density $\sigma_F(x)$ is independent of the underlying distribution $q(x)$, is solely characterized by $F$, and therefore generalizes Wigner's semicircle distribution $\sigma_W(x)$. We find that the level-spacing distributions $p(s)$ are independent of $q(x)$, are dependent upon the inner-product $F$ and whether $A$ is real or complex, and therefore generalize the Wigner's surmise for level spacing. Our results suggest $F$-dependent generalizations of the well-known Gaussian Orthogonal Ensemble (GOE) and Gaussian Unitary Ensemble (GUE) classes. 
\end{abstract}
\maketitle

\section{Introduction}
\label{sec:intro}
Since their beginning in the field of nuclear physics in 1950s, statistical properties of random matrices with specific symmetries have been a source of ongoing investigations~\cite{wigner,ginibre,dyson,mehta}. Wigner's semicircle law for the eigenvalue density, and his surmise for the eigenvalue spacings are some of the most inspired results in random matrix theory. For real, symmetric, $N\times N$ matrices with independent, identically distributed (i.i.d.) entries drawn from an arbitrary distribution $q(x)$ that has zero mean, variance one, and finite higher moments, Wigner showed that the level density in the large-$N$ limit is given by $\sigma_W(x)=(2/\pi)\sqrt{1-x^2}$. He also surmised that the level-spacing distribution for random, real, symmetric matrices, $p^{GOE}(s)=(\pi s/2)\exp(-\pi s^2/4)$,  and for random, complex, Hermitian matrices, $p^{GUE}(s)=(32 s^2/\pi^2)\exp(-4 s^2/\pi)$, are independent of the underlying probability distribution $q(x)$. The near-universality of these results led to the Gaussian Orthogonal Ensamble (GOE) conjecture for random, real, symmetric matrices, the Gaussian Unitary Ensamble (GUE) conjecture for random, complex, Hermitian matrices, and the Gaussian Symplectic Ensamble (GSE) conjecture for fermionic Hamiltonians. The tremendous analytical progress in this field is based on these conjectures which imply that it is sufficient to consider Gaussian distributed entries for the random matrices~\cite{mehta}. We note that the proof of Wigner's semicircle law for an arbitrary, non-Gaussian distribution $q(x)$ is based on the moment method~\cite{wigner}. We also emphasize that the level-spacing distribution $p^{GOE}(s)$, although not exact, is an excellent approximation to the exact answer obtained for the GOE, and it is believed that the result is valid for random, real, symmetric matrices with arbitrary (non-Gaussian) underlying probability distributions $q(x)$~\cite{mehta}.  

These ensembles arise from a fundamental constraint: the Hermiticity of the random Hamiltonian, that guarantees real eigenvalues, and eigenvectors that are orthogonal with respect to the standard inner-product in quantum theory~\cite{mehta}. However, a large class of non-Hermitian matrices - parity and time-reversal symmetric Hamiltonians~\cite{bender,znojil,mark}, rate-equation matrices~\cite{timm}, central potentials in momentum space~\cite{bill}, etc. - has real spectra although the eigenvectors are not orthogonal under the standard inner-product~\cite{bender}.   

This observation raises the following question:  what are the properties of random matrices that are self-adjoint with respect to a {\it general inner product} $F=F^\dagger$ and therefore have purely real eigenvalues? A general inner-product $F$, which, by definition, is a positive-definite matrix, may represent a system (with non-Euclidean geometry) that is not translationally invariant. For example, consider a finite disk with rotational symmetry around its center. The disorder potential $V(\br)$ at a point $\br$ on such a disk will only depend on the radial distance $r$ from its center, but not on its angular orientation. Such a disorder can result from sputtering deposition (for electrons) or from a circular patterned grating (for light). The resulting momentum-space Hamiltonian for such a potential is not Hermitian; but it has purely real eigenvalues~\cite{bill}. Motivated by the generalization of this example to $D$-dimensions, in this paper, we only consider a diagonal inner product $F:V_N\times V_N\rightarrow\mathbb{C}$ defined by 
\begin{equation}
\label{eq:f}
\langle\phi|\psi\rangle_F =\phi^\dagger F\psi =\sum_{i=1}^{N} f_i\phi^*_i\psi_i 
\end{equation}
where $V_N$ is an $N$-dimensional vector space and $F_{jk}=\delta_{jk}f_j=\delta_{jk} j^{D-1}$. Note that an arbitrary inner-product $\langle\cdot|\cdot\rangle_F$ is invariant under transformations $T:V_N\rightarrow V_N$ that obey $T^\dagger F T=F$. When $F=1$, the standard inner-product, this group of transformations corresponds to the unitary (orthogonal) group over a complex (real) vector space $V$. Under such an inner-product, an operator $M$ is self-adjoint if and only if $M=A F$ where $A$ is Hermitian. In this paper, we numerically and analytically investigate the level density and level-spacing distributions of such matrices $M$, or equivalently $\tilde{M}=\sqrt{F}A\sqrt{F}=\tilde{M}^\dagger$, obtained from random, Hermitian matrices $A$ with i.i.d. entries drawn from an arbitrary probability distribution $q(x)$. 

Our three salient results are as follows: i) The level-density $\sigma_F(x)$ is independent of $q(x)$, is characterized by the inner-product $F$, and therefore, is not the same as Wigner's semicircle distribution, $\sigma_F(x)\neq\sigma_W(x)$. ii) For {\it bulk} eigenvalues, we find that the level-spacing distribution is independent of $q(x)$, and is solely characterized by $F$ and whether the matrix $A$ is real or complex. Since these results are invariant under the group of transformations mentioned above, and depend only on whether $A$ is real or complex, we use $p_F^{GOE}(s)$ and $p_F^{GUE}(s)$ to represent the level-spacing distributions. Thus $p_F^{GOE}(s)$ and $p_F^{GUE}(s)$ provide $F$-dependent generalizations of the GOE and GUE universality classes for real and complex random matrices respectively. iii) In each case, as $D$ increases, the level-spacing distribution shifts towards the origin and decays slowly compared to the GOE and GUE results. 

The plan of the paper is as follows. In Sec.~\ref{sec:ld} we present numerical data for the level density which show a clear $F$-dependence, and discuss the properties of the level density $\sigma_F(x)$. Section~\ref{sec:ls} starts with numerical data for level-spacing distributions in the real and complex cases. We then present a simple model that qualitatively explains the evolution of level-spacing distributions with increasing $D\geq 1$ as the inner-product $F_D$ deviates from the standard-inner product, $F_{D=1}=1$. We conclude the paper in Sec.~\ref{sec:disc} with a brief discussion. Although the non-Hermitian, random matrices $M$ considered in this paper appear similar to those in Refs.~\cite{shukla,gg,wish,pandey}, we point out the crucial differences among them in Sec.~\ref{sec:disc}. We emphasize that the probability distribution for the random matrices considered here is $p(M)\propto\exp(-\Tr(A^\dagger A))\propto\exp(-\Tr(F^{-2}M^\dagger M))$; it is not a function of $\Tr(M^\dagger M)$, and therefore, the traditional methods~\cite{mehta,gg,pastur,brezin} used to analyze eigenvalue statistics may not be applicable.

 \section{Level Density}
\label{sec:ld} 
We start this section with numerical data that hint at our result. For independent entries of the random matrix $A$, we use a Gaussian distribution $G(x)$ with zero mean and variance one, or a uniform distribution $U(x)$ with zero mean and variance one, or a distribution $q_\theta(x)$ that corresponds to random entries $r_i=\cos\theta G_i +\sin\theta U_i$. Note that the distribution $q_\theta(x)=q_\theta(-x)$ interpolates continuously from the Gaussian ($\theta=0$)  to the uniform ($\theta=\pi/2$), has zero mean, variance one, and finite higher moments. Since the variance of i.i.d. entries is one, the eigenvalue scale for the level density is given by $\Lambda_F=2\Tr F/\sqrt{N}$ ($\Lambda_F=2\sqrt{2}\Tr F/\sqrt{N}$) when $M$ is a real (complex) matrix. This scale implies that the second moment $\mu_2$ of the resulting level density is fixed, $\mu_2=1/4$~\cite{brezin,miller}. 
\begin{figure}[thb]
\begin{center}
\includegraphics[angle=0,width=9cm]{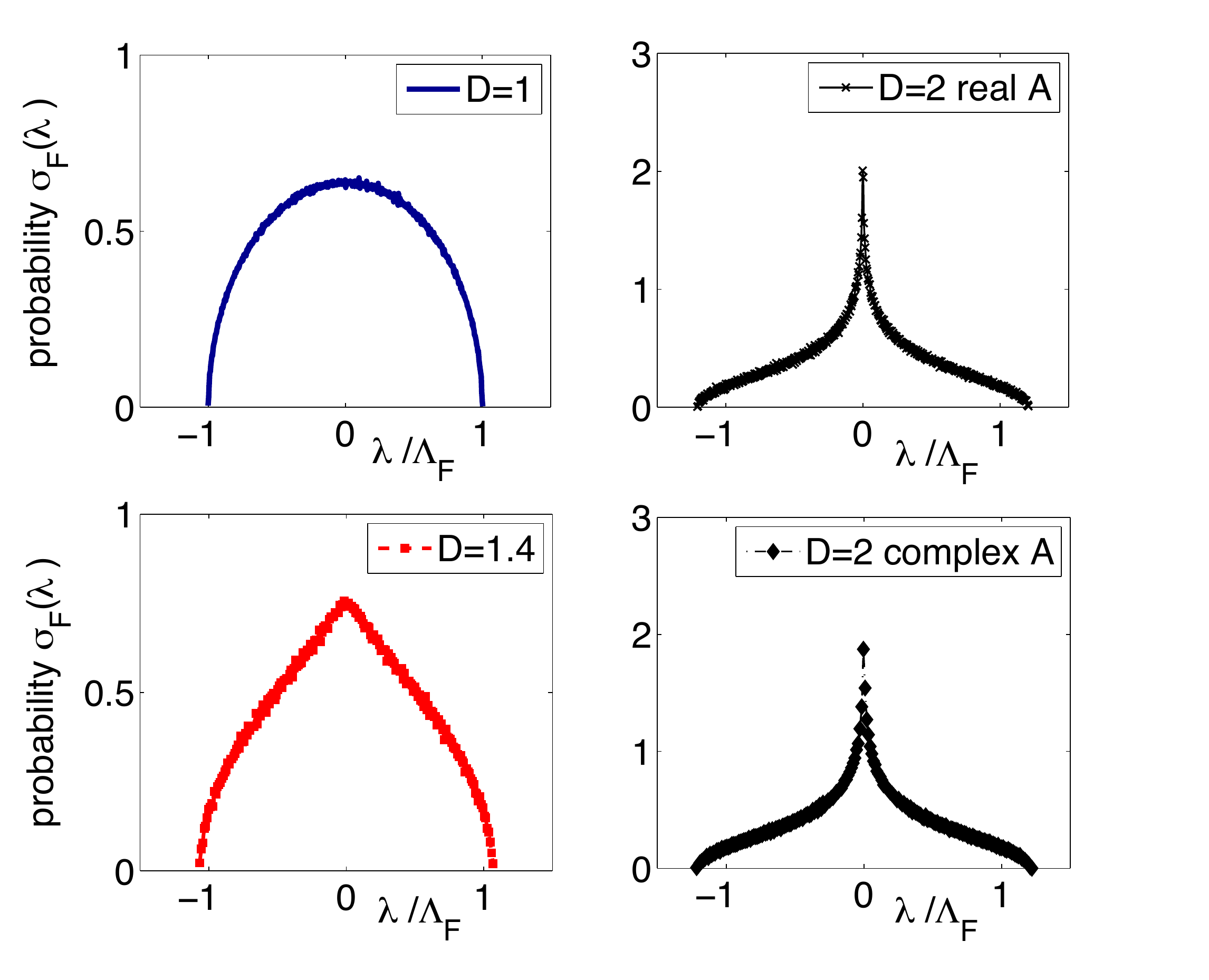}
\caption{\label{fig:sigmafd}
(color online) Level-density $\sigma_F(x)$ for $D=\{1,1.4,2\}$. The top-left panel reproduces Wigner's result for $D=1$, obtained using $N_t=10^3$, $N=3\times 10^3$, $N_b=500$, and Gaussian distributed complex entries ($\theta=0$). The bottom-left panel shows the result for $D=1.4$ with $N_t=500$, $N=10^3$, $N_b=200$, and real entries with $\theta=\pi/4$. The top-right panel shows the result for $D=2$, $N_t=500$, $N=2\times 10^3$, $N_b=200$, and real entries from a uniform distribution ($\theta=\pi/2$), whereas the bottom-right panel shows the result for the same $D$, with $N_t=10^3$, $N=3\times 10^3$, $N_b=500$ and complex entries from a distribution $q_\theta(x)$ with $\theta=\pi/3$. As $D$ increases, we see that the level-density $\sigma_F(x)$ sharpens near the origin, broadens, and, when scaled appropriately, is the same for both real and complex random matrices.}
\end{center}
\end{figure}

Figure~\ref{fig:sigmafd} shows the level density for $D=\{1, 1.4, 2\}$. We have verified that these results are essentially independent of the number of trials $N_t\gg 1$, the matrix size $N\geq 10^2$, the number of bins $N_b$, and the underlying probability distribution $q_\theta(x)$. The top-left panel in Fig.~\ref{fig:sigmafd} reproduces Wigner's result $\sigma_W(x)$ that is expected when $D=1$. The bottom-left panel ($D=1.4$) and the top-right panel ($D=2$) show that with increasing $D$, the level density $\sigma_F(x)$ broadens and becomes sharply peaked near the origin. The top-right and bottom-right panels show that when scaled appropriately, the level density $\sigma_F(x)$ is the same for real and complex matrices and thus depends only on the inner-product $F$.  These numerical data strongly suggest that $\sigma_F(x)$ is independent of $q_\theta(x)$ and is solely characterized by the inner product $F$. 

\begin{figure}[t!]
\begin{center}
\includegraphics[angle=0,width=9cm]{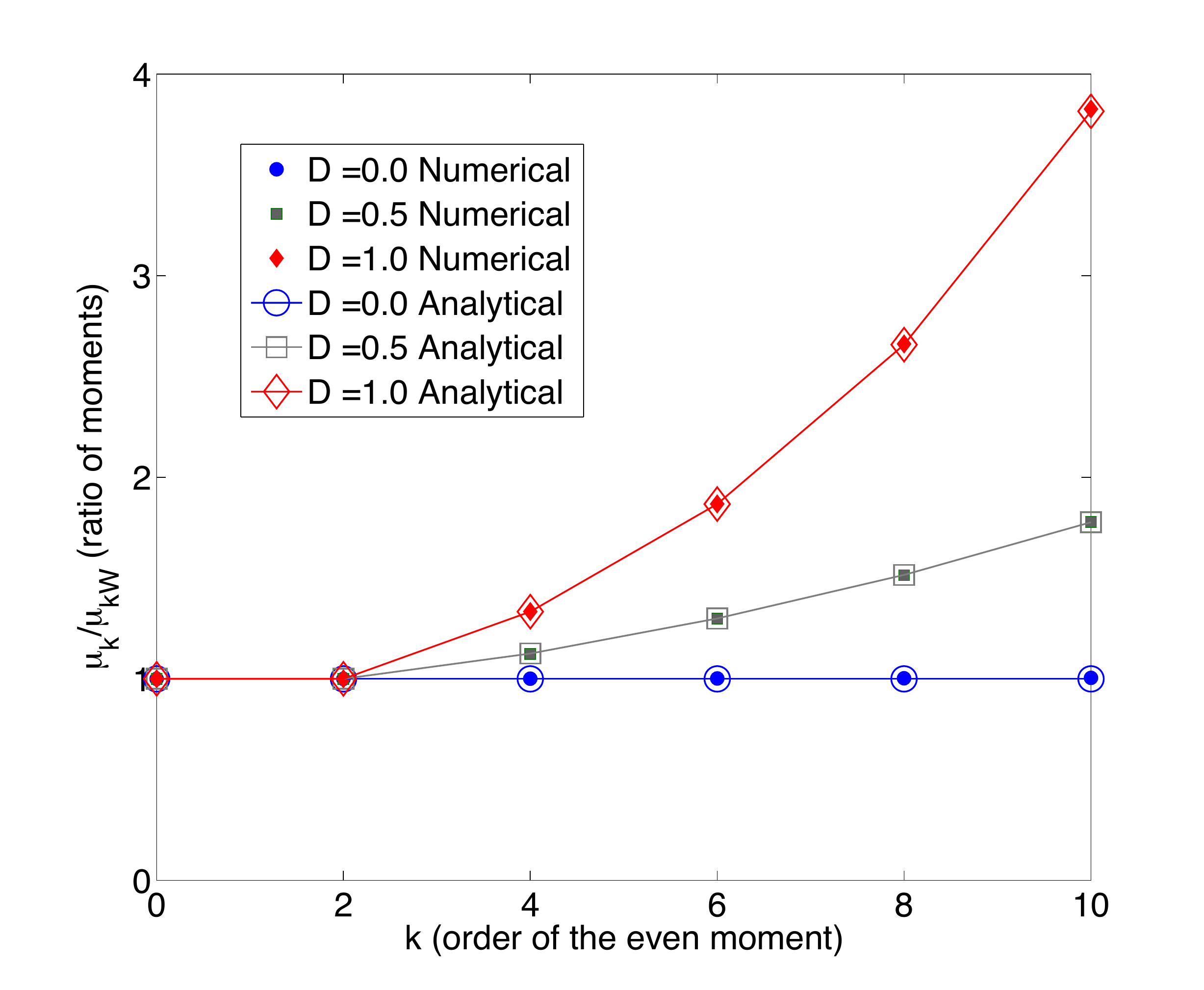}
\caption{\label{fig:evenmoments}
(color online) Even moments $\mu_k$ of the level density $\sigma_F(x)$ scaled by their values $\mu_{kW}$ for the Wigner distribution, for $D=\{1,1.5,2\}$. Each numerical data-point (solid symbols) is obtained by using a different set of values for $(N_t, N, q_\theta(x))$. As $D$ increases the ratio $\mu_k/\mu_{kW}\sim D^{(k/2-1)}$ increases, consistent with the broadening of the level density $\sigma_F(x)$ relative to the $D=1$ result.}
\end{center}
\end{figure}

To characterize the $D$-dependence of the resulting level density, we calculate its even moments and compare them with the results for the Wigner distribution $\sigma_W(x)$~\cite{wigner,miller}. The $k^{\mathrm{th}}$ moment ($k$ even) of the level density is given by $\mu_k=\lim_{N\rightarrow\infty}\langle\Tr((AF)^k)/N\rangle_A$ where $\langle\cdots\rangle_A$ denotes random averaging. It is straightforward, but tedious, to calculate the non-vanishing averages systematically; for $k=\{4,6,8\}$ the results are 
\begin{eqnarray}
\label{eq:moment4}
\mu_{4} & = & \lim_{N \rightarrow \infty} \frac{N}{2^4} \left[ 2\frac{\Tr F^2}{( \Tr F )^2} \right],\\
\label{eq:moment6}
\mu_{6} & = & \lim_{N \rightarrow \infty} \frac{N^2}{2^6} \left[ 3\frac{(\Tr F^2)^2}{( \Tr F )^4} + 2 \frac{ \Tr F^3}{( \Tr F )^3} \right],\\
\label{eq:moment8}
\mu_{8} & = & \lim_{N \rightarrow \infty} \frac{N^3}{2^8}\left[ 8\frac{\Tr F^2\Tr F^3}{(\Tr F)^5}+4\frac{(\Tr F^2)^3}{(\Tr F)^6}+2 \frac{\Tr F^4}{(\Tr F)^4}\right].
\end{eqnarray}
Note that when $F=1$ the sum of terms in a square bracket gives the corresponding Catalan number~\cite{wigner,miller}. Since $F={\rm diag}(j^{D-1})$, $\Tr F^m\sim N^{m(D-1)+1}$ for $D>0$ and finite moments require $D\geq 1$. Fig.~\ref{fig:evenmoments} shows the numerically obtained even moments scaled by their values for $D=1$, $\mu_k/\mu_{kW}$ (solid symbols), and the analytical results (open symbols) for $D=\{1,1.5,2\}$. The monotonic increase in the even moments with $D$ is consistent with the broadening of the level density $\sigma_F(x)$ relative to the Wigner's result $\sigma_W(x)$. Although an exact solution for $\sigma_F(x)$ is known through implicit equations~\cite{baizhang}, our (numerical and analytical) approach provides detailed information. 

We recall that the level-density $\sigma_F(x)$ is independent of the underlying probablity distribution $q(x)$ only when $q(x)$ has finite higher moments; for example, even for $F=1$, if $q(x)$ is the Cauchy distribution, the resulting level density is not the same as the Wigner distribution $\sigma_W(x)$~\cite{miller}. We also point out that the existence of the asymptotic limit in Eqs.(\ref{eq:moment4})-(\ref{eq:moment8}) is determined by the large-$N$ behavior of the inner-product elements $F_{jk}=\delta_{jk}f_j$. In particular, if $\lim_{N\rightarrow\infty}\Tr F$ is finite or diverges as $N^\alpha$ with $\alpha<1$, it follows that the level density $\sigma_F(x)$ has divergent higher moments and therefore it cannot have a compact support. Numerical data obained by using $F_{jk}=\delta_{jk} j^{D-1}$ with $D<1$ or $F_{jk}=\delta_{jk}e^{-j\alpha}$ show that the support of the level density $\sigma_{F,N}(x)$ widens as $N$ increases. Thus, universal results for the level density $\sigma_F(x)$ are obtained only when the underlying probability distribution $q(x)$ and the inner-product $F$ obey the constraints established here.

\section{Level-spacing Distribution} 
\label{sec:ls}

The level spacing distribution $p_F(s)$ is formally given by 
\begin{equation}
\label{eq:lss}
p_F(s)=\lim_{N\rightarrow\infty}\left\langle\frac{1}{N}\sum_{i=1}^{N} \delta\left(s-\frac{\lambda_{i+1}-\lambda_i}{\Delta\lambda_{\mathrm{avg}}}\right)\right\rangle_A
\end{equation}
where $\lambda_i(M=AF)$ are the eigenvalues listed in ascending order, $\lambda_{i+1}\geq\lambda_i$, and $\Delta\lambda_{\mathrm{avg}}=\sum_{j=1}^{N-1}(\lambda_{j+1}-\lambda_j)/(N-1)$ is the average spacing between the eigenvalues. Note that the eigenvalue scale $\Lambda_F$ does not explicitly enter Eq.(\ref{eq:lss}). 

\begin{figure}[t]
\begin{center}
\includegraphics[angle=0,width=9cm]{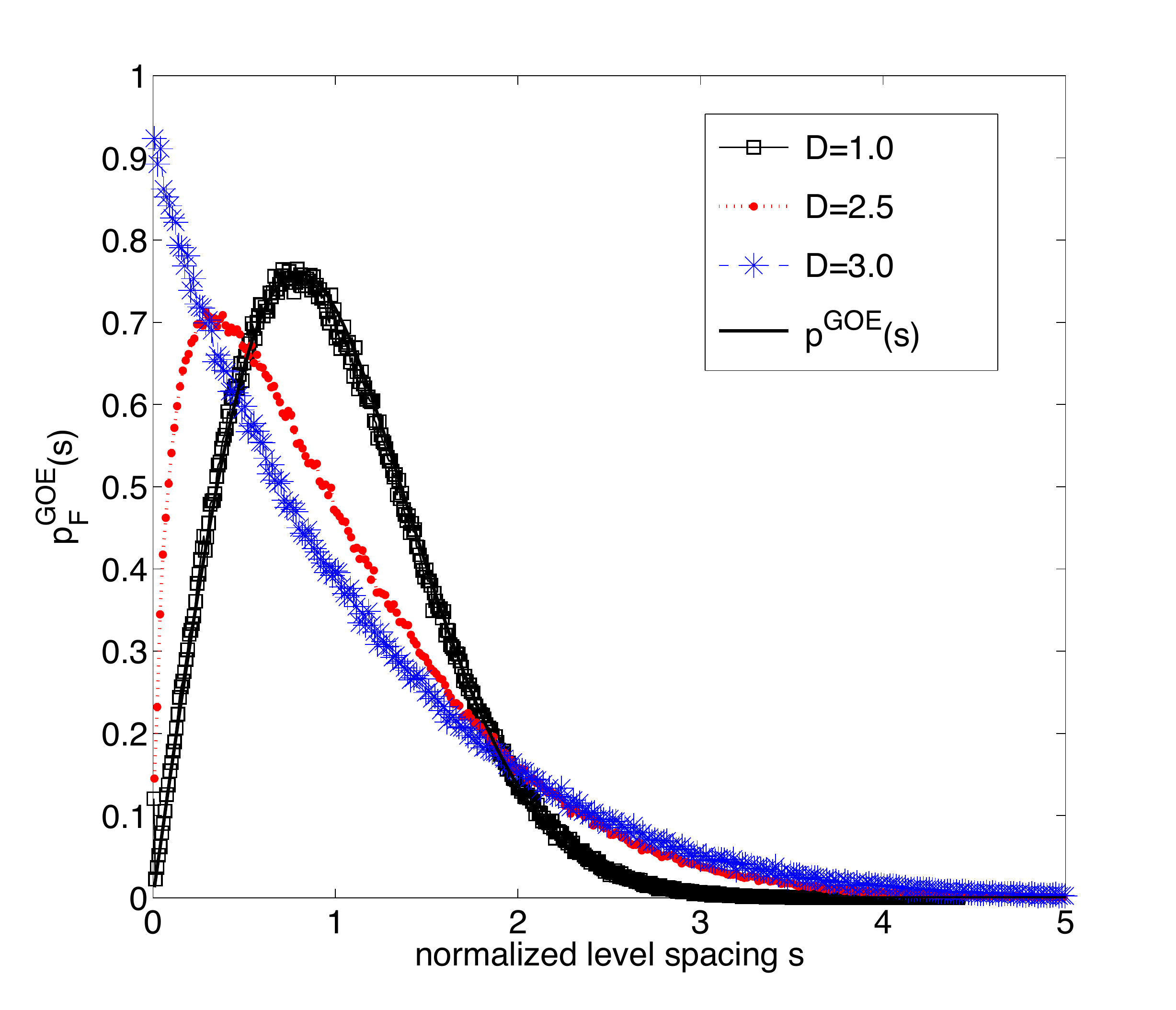}
\caption{\label{fig:lssgoe}
(color online) Level-spacing distribution $p^{GOE}_F(s)$ for random, {\it real}, $N\times N$ matrices $M=AF$ with $D=\{1,2.5, 3\}$. The $D=1$ results (black open squares) are obtained by using $N_t=10^3$, $N=2\times 10^3$, $N_b=500$, and a Gaussian distribution, and match well with Wigner surmise $p^{GOE}(s)$ (black solid line). The $D=2.5$ results (red circles), obtained by using $N_t=10^3$, $N=3\times 10^3$, $N_b=500$, and $\theta=\pi/4$, show that the level-spacing distribution maximum shifts towards the origin, its slope increases, and it broadens. The $D=3$ results (blue stars) are obtained by using $N_t=100$, $N=10^4$, $N_b=500$, and a uniform distribution.}
\end{center}
\end{figure}
Figure~\ref{fig:lssgoe} shows numerically obtained level-spacing distribution for the bulk ($\geq 10\%$) eigenvalues of random, real, $N\times N$ matrices $M=AF$ for $D=\{1, 2.5, 3\}$. Note that the level-spacing scale is chosen so that the average level-spacing is one: $\int_0^{\infty} s p_F(s)ds=1$. We have verified that these results are independent of $N_t\gg 1$, $N\geq 10^2$, and  $q_\theta(x)$ used to create the matrix $A$. When $D=1$, $F$ is the identity matrix and the numerical results (black squares) match well with Wigner surmise $p^{GOE}(s)$ (black solid line). As $D$ increases, we see that the level-spacing distribution becomes broader, the weight of the distribution near the origin increases and so does its slope at the origin. We recall that this  distribution is invariant under real matrix transformations $O$ that obey $O^{T}F O=F$. Since when $F=1$, they correspond to the group of orthogonal matrices, we use the notation $p^{GOE}_F(s)$ to denote the level spacing results that are invariant under the $F$-dependent group of real matrix transformations. 

\begin{figure}[t]
\begin{center}
\includegraphics[angle=0,width=9cm]{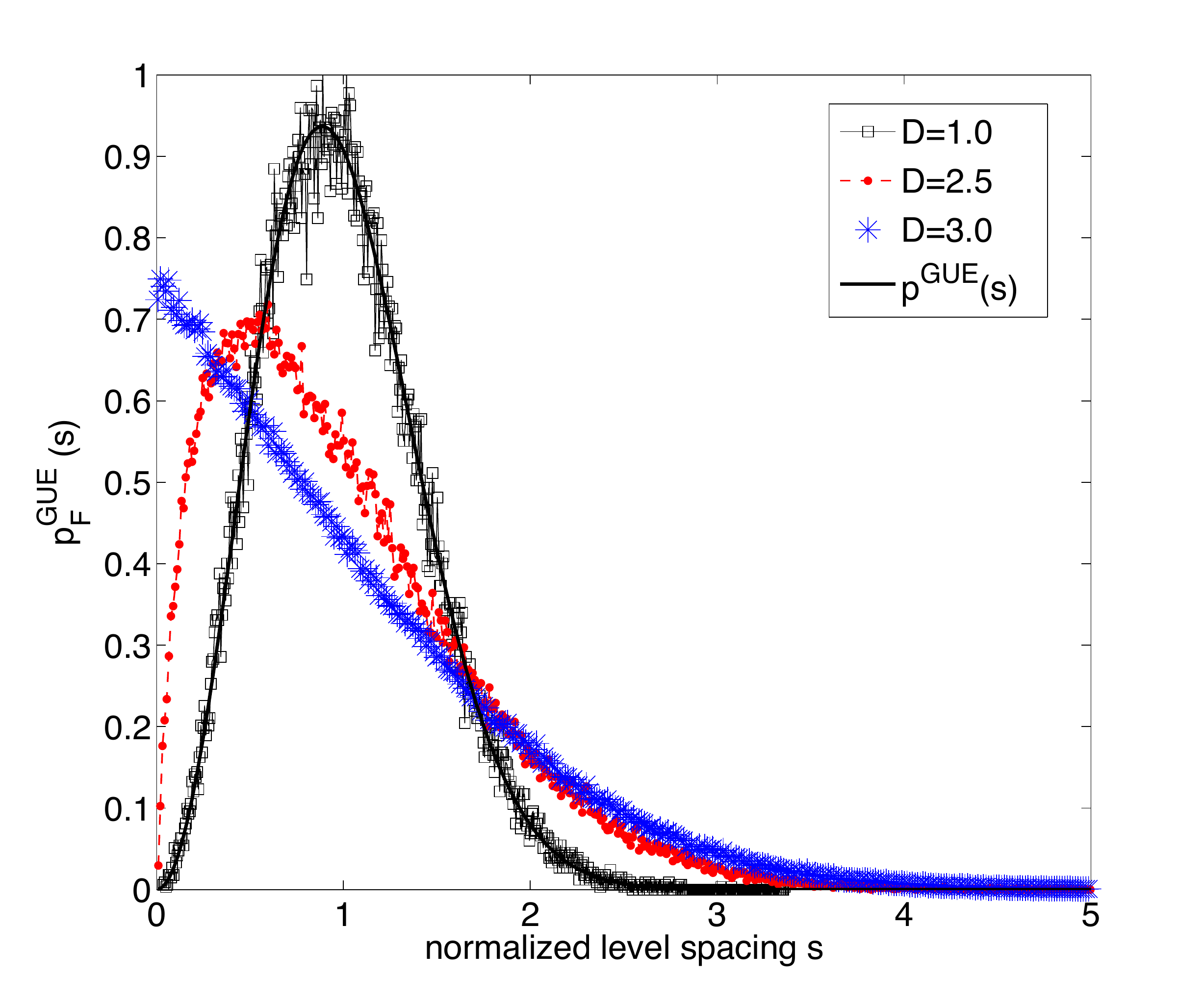}
\caption{\label{fig:lssgue}
(color online) Level-spacing distribution $p^{GUE}_F(s)$ for random, {\it complex}, $N\times N$ matrices $M=AF$ with $D=\{1, 2.5, 3\}$. The $D=1$ results (black open squares) are obtained by using $N_t=10^3=N$, $N_b=200$, and a Gaussian distribution, and match well with the Wigner's surmise $p^{GUE}(s)$ (black solid line). The $D=2.5$ results (red circles), obtained using $N_t=10^3=N$, $N_b=500$, and $\theta=\pi/3$, show that the level-spacing distribution maximum shifts towards the origin, its second derivative at the origin increases, and it broadens. The $D=3$ results (blue stars) are obtained by using $N_t=100$, $N=3\times10^3$, $N_b=500$, and a uniform distribution.}
\end{center}
\end{figure}
Figure~\ref{fig:lssgue} shows the level-spacing distribution for the bulk ($\geq 10\%$) eigenvalues of random, complex, $N\times N$ matrices $M=AF$ for $D=\{1, 2.5, 3.0\}$. When $D=1$, the numerical results (black squares) match with the Wigner's surmise $p^{GUE}(s)$ (black solid line). As $D$ increases, we see that the level-spacing distribution becomes broader, the weight of the distribution near the origin increases and so does its second derivative at the origin. This distribution is invariant under complex matrix transformations $U$ that obey $U^{\dagger}F U=F$. Since when $F=1$, they correspond to the group of unitary matrices, we use the notation $p^{GUE}_F(s)$ to denote the level spacing results that are invariant under the $F$-dependent group of complex matrix transformations. These numerical results strongly suggest that the level-spacing distributions $p^{GOE}_F(s)$ and $p^{GUE}_F(s)$ are independent of the underlying probability distribution $q_\theta(x)$ and are $F$-dependent generalizations of the GOE and GUE universality classes. 

The $D$-dependence of the large-$N$ level-spacing distributions may be qualitatively understood by exploring it for $2\times 2$ matrices $M=AF$. Here, $A$ is a $2\times 2$ random, real symmetric or complex Hermitian matrix with entries drawn from a Gaussian distribution with zero mean, variance $\sigma$, and $F={\rm diag}(1,f_2)$ where we choose $f_2=N^{D-1}\gg 1$ to mimic the effects of a non-standard inner-product $F$. We remind the Reader that when $D=1$, this procedure gives excellent approximations, $p^{GOE}(s)$ and $p^{GUE}(s)$,  to the large-$N$ results~\cite{wigner,herman}. When $D>1$, for real, $2\times 2$ matrices $M=AF$, the result is~\cite{isaker} 
\begin{equation}
\label{eq:goe}
p^{GOE}_{2\times 2}(s)=\frac{4}{\alpha\beta} s e^{-B_{+}s^2}I_0(B_{-}s^2)
\end{equation}
where $\alpha^2=16\sigma^2f_2$, $\beta^2=8\sigma^2(1+f_2^2)$, $B_\pm=(\alpha^{-2}\pm\beta^{-2})$, and $I_0(x)$ is the modified Bessel function of first kind. The variance $\sigma$ is chosen so that the mean level-spacing is unity. In the limit $f_2=N^{D-1}\gg 1$, $B_\pm\rightarrow \pi f_2/32$, $(B_{+}-B_{-})\rightarrow 1/\pi$, and the constraint on the mean-level spacing leads to $p^{GOE}_{2\times 2}(s)=(\pi\sqrt{f_2/32}) s\exp(-s^2/\pi)$. Thus, as $D$ increases, the distribution $p^{GOE}_{2\times 2}(s)$ shifts towards the origin, the slope of the distribution near the origin diverges, and at large $s$, the distribution decays more slowly than $p^{GOE}\sim\exp(-\pi s^2/4)$. When the $2\times 2$ matrix $A$ is complex, the result for the level-spacing distribution is given by  
\begin{equation}
\label{eq:gue}
p^{GUE}_{2\times 2}(s)=\sqrt{\frac{2}{\pi}}\mu^2\nu s^2 e^{-\nu^2 s^2/2}\frac{D(s\tau)}{s\tau}
\end{equation}
where $\mu^{-2}=4\sigma^2 f_2$, $\nu^{-2}=2\sigma^2(1+f_2^2)$, $D(x)$ is the Dawson's integral which satisfies $D(x)\sim x$ when $x\ll 1$ and $D(x)\sim 1/(2x)$ when $x\gg 1$, and $\tau^2=(\mu^2-\nu^2)/2$. When $f_2=N^{D-1}\gg 1$, the constraint on the mean-level spacing implies that $\sigma f_2\sqrt{4/\pi}=1$. This leads to $p^{GUE}_{2\times 2}(s)=\sqrt{8f_2/\pi^3} s e^{-s^2/\pi} D(s\sqrt{f_2/2\pi})$. Therefore, as $D>1$ increases, the distribution $p^{GUE}_{2\times 2}(s)$ shifts towards the origin, the second derivative of the distribution at the origin diverges, and at large $s$ the distribution decays more slowly than $p^{GUE}\sim\exp(-4 s^2/\pi)$. 

\begin{figure}[t!]
\begin{center}
\includegraphics[angle=0,width=9cm]{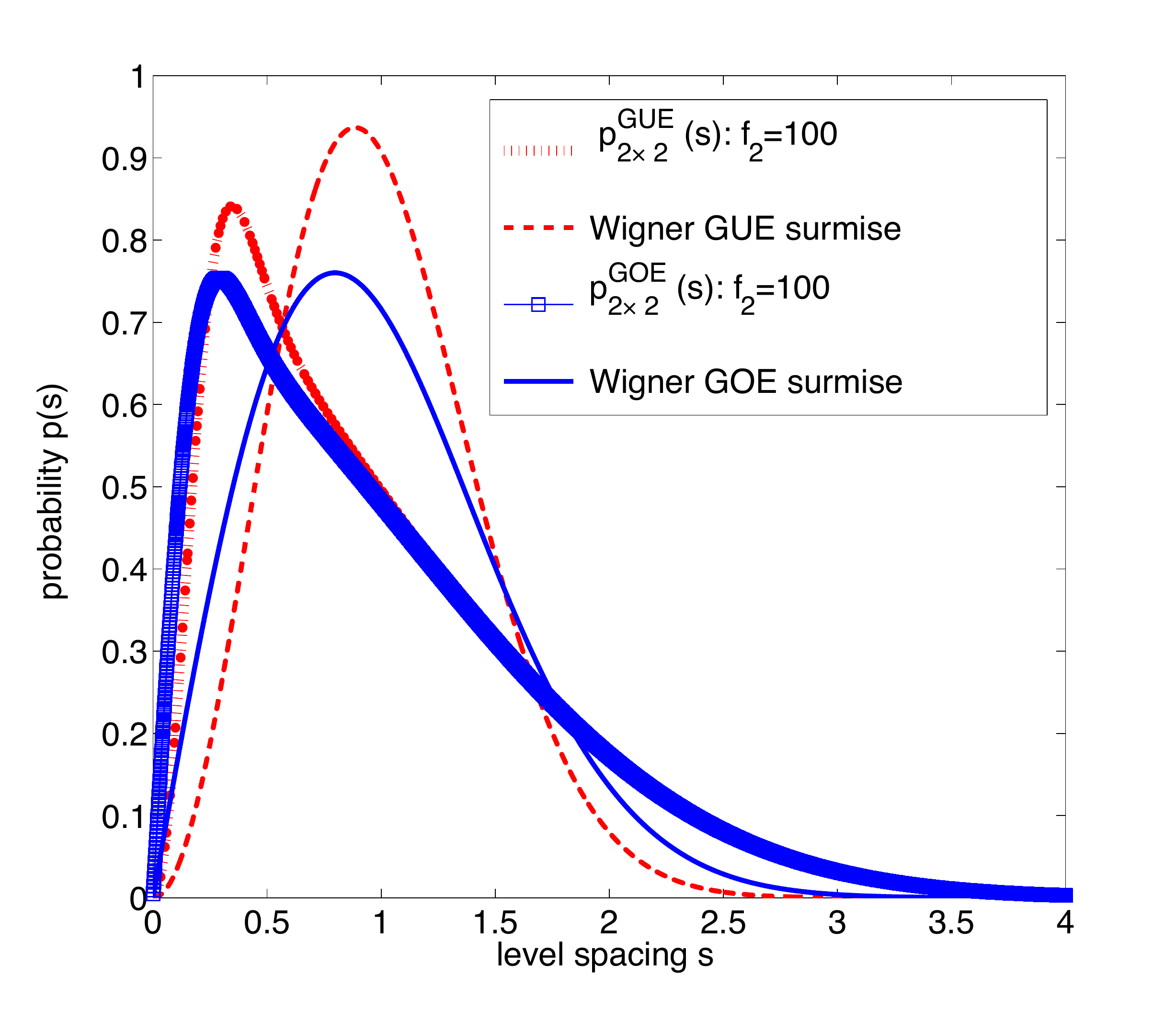}
\caption{\label{fig:exact}
(color online) Normalized level-spacing distribution $p^{GOE}_{2\times 2}$, Eq.(\ref{eq:goe}), (blue squares thick line), the Wigner surmise $p^{GOE}$ (blue solid thin line), the distribution $p^{GUE}_{2\times 2}$, Eq.(\ref{eq:gue}), (red dotted line), and the Wigner surmise $p^{GUE}$ (red dashed line).  In each case, as $f_2\gg 1$ increases, the distribution $p^{GOE}_{2\times 2}$ ($p^{GUE}_{2\times 2}$) shifts towards the origin, its slope (second derivative) at the origin increases, and the distribution broadens.}
\end{center}
\end{figure}

Figure~\ref{fig:exact} shows the level-spacing distributions $p^{GOE}_{2\times 2}(s)$ (blue squares thick line), the corresponding Wigner surmise $p^{GOE}(s)=(\pi s/2)\exp(-\pi s^2/4)$ (blue solid thin line), the distribution for complex matrices $p^{GUE}_{2\times 2}(s)$ (red dotted line), and the corresponding Wigner surmise $p^{GUE}(s)=(32 s^2/\pi^2)\exp(-4 s^2/\pi)$ (red dashed line) obtained by using $f_2=100\gg 1$. We emphasize that although the $2\times 2$-case analytical results {\it qualitatively correspond} to the large-$N$ numerical data in regions $s\ll 1$ and $s>1$, they are not a good approximation in the intermediate range, and the analytical answer for the large-$N$ level-spacing distributions $p^{GOE}_F(s)$ and $p^{GUE}_F(s)$ remains unknown. 

\section{Discussion} 
\label{sec:disc}
In this paper, we have investigated the level-density and level-spacing distributions for matrices $M=AF\neq M^\dagger$ that are self-adjoint with respect to a diagonal inner-product $F$, motivated by the radial integral in $D\geq 1$ dimensions. 

We have shown that the level density $\sigma_F(x)$ is dependent upon, and solely characterized by the inner-product $F$. With increasing $D$, the level density $\sigma_F(x)$ develops a peak at the origin, and broadens its base, but continues to have a compact support.  We found that the level-spacing distribution $p_F(s)$ for such matrices is dependent upon the inner-product $F$, and upon whether the matrix $A$ is real or complex. As $D$ increases, the level-spacing distributions $p^{GOE}_F(s)$ and $p^{GUE}_F(s)$ shift towards the origin, become steeper near the origin, and broaden. Our numerical results and qualitative analysis thus strongly suggest $F$-dependent generalizations of the GOE and GUE universality classes; in each case, however, analytical expressions for the $F$-dependent $n$-point correlation functions remain unknown~\cite{mehta}. 

The analysis in this paper is limited to a specific functional form of the inner-product $F$. Our results, however, raise the broader question of universality classes for random matrices that are self-adjoint with respect to a generic, non-diagonal, positive-definite, inner product $F$. The random, non-Hermitian matrices $M=AF$ are similar to those in Refs.~\cite{shukla,gg} but for the crucial difference that, in contrast to Refs.~\cite{shukla,gg}, the eigenvalues of $M=AF$ are {\it always purely real}. The probability distribution $p(M)\propto\exp(-\Tr(A^\dagger A))$ is also similar to the measure for random, generalized Wishart matrices~\cite{wish,pandey}. In contrast to Refs.~\cite{wish,pandey}, however, we have explored the eigenvalues of the matrix $M=AF$, and not the positive-definite matrix $A^\dagger A$. The generalizations of Wigner's semicircle law and the GOE and GUE surmises for level-spacing distributions, presented in this paper, suggest that the statistical properties of random, self-adjoint matrices, although independent of the underlying probability distribution $q(x)$, are richly varied. 


\begin{acknowledgements}
W.A.K. was supported by a IUPUI Undergraduate Research Opportunities Program (UROP) grant. 
\end{acknowledgements}


\end{document}